\documentclass[preprint,3p,times,twocolumn]{elsarticle}

\usepackage{prletters}

\usepackage{amssymb}
\usepackage{amsmath}

\begin{document}

\begin{frontmatter}



\title{A dark and bright channel prior guided deep network for retinal image quality assessment}



\author[a]{Ziwen Xu}

\author[a,b]{Beiji Zou}

\author[a]{Qing Liu\corref{mycorrespondingauthor}}
\cortext[mycorrespondingauthor]{Corresponding author}
\ead{qing.liu.411@gmail.com}

\address[a]{School of Computer Science and Engineering, Central South University, Changsha 410083, China}
\address[b]{Hunan Province Engineering Technology Research Center of Computer Vision and Intelligent Medical Treatment, Changsha 410083, China}




\begin{abstract}
Retinal image quality assessment is an essential task in the diagnosis of retinal diseases. Recently, there are emerging deep models to grade quality of retinal images. Current state-of-the-arts either directly transfer classification networks originally designed for natural images to quality classification of retinal images or introduce extra image quality priors via multiple CNN branches or independent CNNs. This paper proposes a dark and bright channel prior guided deep network for retinal image quality assessment called GuidedNet. Specifically, the dark and bright channel priors are embedded into the start layer of network to improve the discriminate ability of deep features. In addition, we re-annotate a new retinal image quality dataset called RIQA-RFMiD for further validation. Experimental results on a public retinal image quality dataset Eye-Quality and our re-annotated dataset RIQA-RFMiD demonstrate the effectiveness of the proposed GuidedNet.
\end{abstract}



\begin{keyword}
Retinal image quality assessment        
\newline Deep network 
\newline Dark channel prior
\newline Bright channel prior



\end{keyword}

\end{frontmatter}


\section{Introduction}
\label{sec:intro}
High-quality retinal images are required for the diagnosis of diabetic retinopathy, glaucoma and other retinal disorders \cite{quellec2020automatic, shankar2020automated}. They facilitate ophthalmologists make correct clinical decisions efficiently. On the contrary, low-quality images may confuse ophthalmologists. What is worse, for computer-aided retinal image analysis systems which commonly are designed with high-quality retinal images, low quality retinal images would be a catastrophe. Thus it would be desired to automaticaly assess the retinal image quality and filter low-quality images before performing downstream tasks. Clinically, retinal image quality assessment (Retinal-IQA) is performed by a well trained optometrist manually, which heavily depends on operator's experience and is time-consuming. To improve the efficiency of retinal image acquisition, automated Retinal-IQA becomes necessary.
\begin{figure}[htb]
	\centering
	\includegraphics[width=0.88\linewidth]{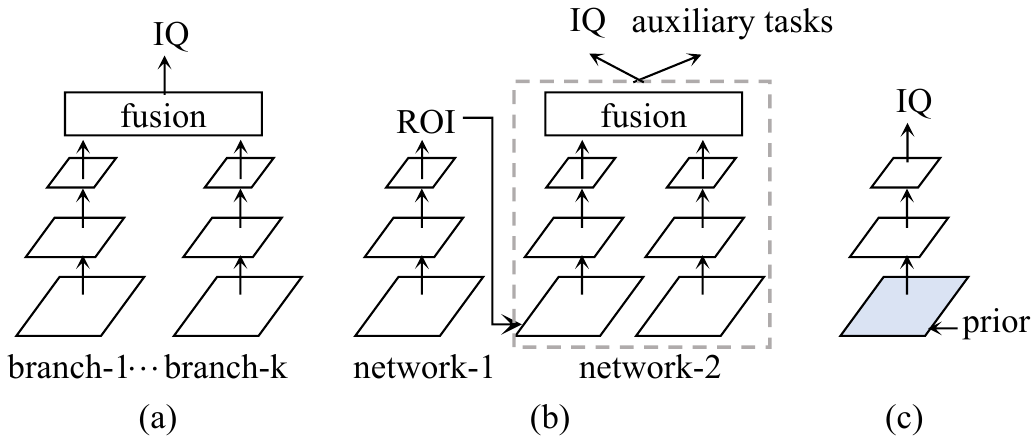}
	\caption{Comparison of different CNN architectures combining priors for Retinal-IQA. (a) Combining priors via multiple branches. (b) Combining priors via multiple independent networks and branches. (c) Our proposed prior guided single branch network.}
	\label{fig:fmcompare}
\end{figure}

The goal of automated image quality assessment (IQA) is to grade images in terms of quality measures. According to type of image, IQA can be subdivided into natural image quality assessment (Natural-IQA) \cite{mittal2012no,ou2019novel,yan2018two}, Retinal-IQA \cite{zago2018retinal,yu2017image}, etc. Most IQA methods are designed for natural images, however, Natural-IQA methods may be not suitable for Retinal-IQA. The reason is that Natural-IQA methods usually rely on the assumption that high-quality natural images have certain statistics which can be damaged by low-quality factors. However, these statistics are not consistent with those of retinal images. Current state-of-the-art Retinal-IQA methods solve it with modern CNNs as they have achieved huge successes in the field of computer vision. Some works such as \cite{zago2018retinal} and \cite{yu2017image} directly fine-tune CNNs originally designed for natural scene image classification with retinal images to learn rich deep features for Retinal-IQA. To exploit more powerful representation for Retinal-IQA, complex frameworks involving multiple parallel CNN branches as shown in Fig. \ref{fig:fmcompare}(a) or multiple independent CNN networks as shown in Fig. \ref{fig:fmcompare}(b) to make use of priors about IQA are proposed. For example, Fu et al. claim that different colour spaces represent different characteristics and propose Multiple Colour-space Fusion Network (MCF-Net) \cite{fu2019evaluation}. MCF-Net unifies three parallel CNN branches into one framework to learn complementary informative contexts from RGB, HSV and Lab colour spaces for retinal-IQA. Muddamsetty et al. propose a combined model for Retinal-IQA \cite{muddamsetty2020multi}, which ensembles deep features from a CNN branch and generic features from two texture branches. In \cite{shen2018multi}, Shen et al. claim that auxiliary tasks contribute to IQA and propose a multi-task framework for Retinal-IQA, named MFIQA. MFIQA \cite{shen2018multi} consists of a ResNet-50 network \cite{he2016deep} for detection of region of interest (ROI), a VGG-16 network \cite{simonyan2014very} for refinement of ROI location, a network consisting of two encoders to encode the global retinal images and local ROI for Retinal-IQA and three auxiliary classification tasks, i.e., artifact, clarity and field definition. It is further improved by introducing domain invariance and interpretability in \cite{shen2020domain}. Comparing to framework consisting of single networks with single branch, those complex frameworks make use of extra priors about IQA and achieve superior performances. However, the parameters to be optimised rapidly multiply. Additionally, in \cite{shen2018multi, shen2020domain}, auxiliary task learning requires extra annotated data and the whole framework can not be trained end-to-end. These motivate us to develop a framework which can make priors incorporate in CNNs without increasing extra parameters and annotated data, as shown in Fig. \ref{fig:fmcompare}(c).
\begin{figure}[htb]
	\centering
	\includegraphics[width=0.82\linewidth]{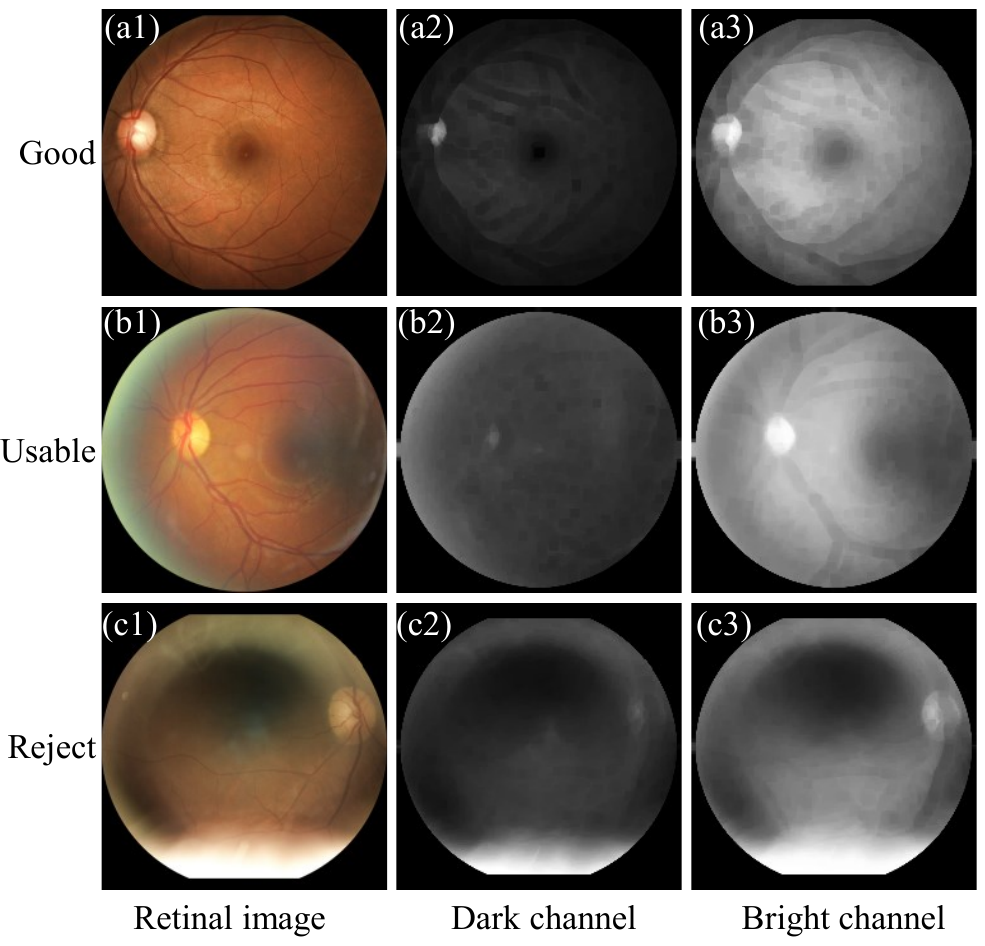}
	\caption{Examples for dark and bright channel priors of different quality retinal images.}
	\label{fig:DARK-BRIGHT}
\end{figure}

To this end, we go back to the basics and notice that, within the context of Retinal-IQA, a high-quality retinal image commonly captured under even illumination, in which salient structures such as optic disc and vessels etc. are clearly and definitely visible as shown in the first example in Fig. \ref{fig:DARK-BRIGHT}. Accordingly, we propose two novel priors. The first one is dark channel prior. It is based on the observation that, in retinal images captured with even illumination, most of local patches contain some pixels which have very low intensities in at least one channel. On the contrary, in retinal images captured with uneven illumination, pixels in regions with strong illumination have high intensities in all channels. It is exactly inline with the dark channel prior which is first proposed in \cite{he2010single}. Fig. \ref{fig:DARK-BRIGHT} shows the dark channels of three examples respectively graded as ‘Good’, ‘Usable’ and ‘Reject’. Obviously, except for the bright structure region, i.e., optic disc region, the intensities of pixels in the dark channel map of image graded as ‘Good’ are always low. On the contrary, the intensities of pixels in the dark channel map of images graded as ‘Usable’ and ‘Reject’ are uneven. Particularly, pixels in regions with strong illumination have high intensities. The second prior is bright channel prior. It is based on the observation that most bright pixels have high intensities for retinal image captured with even illumination. On the contrary, for retinal image captured with uneven illumination, intensities of the bright pixels in regions with weak illumination are low. Bright channels of three examples graded as ‘Good’, ‘Usable’ and ‘Reject’ shown in the third column of Fig. \ref{fig:DARK-BRIGHT} illustrate our observation. To make priors of dark channel and bright channel incorporate in CNNs, we develop a novel framework named GuidedNet. Particularly, in GuidedNet, convolution with fixed kernels is plugged in the first layer to estimate the priors of dark channel and bright channel and guide the network to pay attention to bright regions in dark channel prior map and dark regions in bright channel prior map.
\begin{figure*}[htb]
	\centering
	\includegraphics[width=0.85\linewidth]{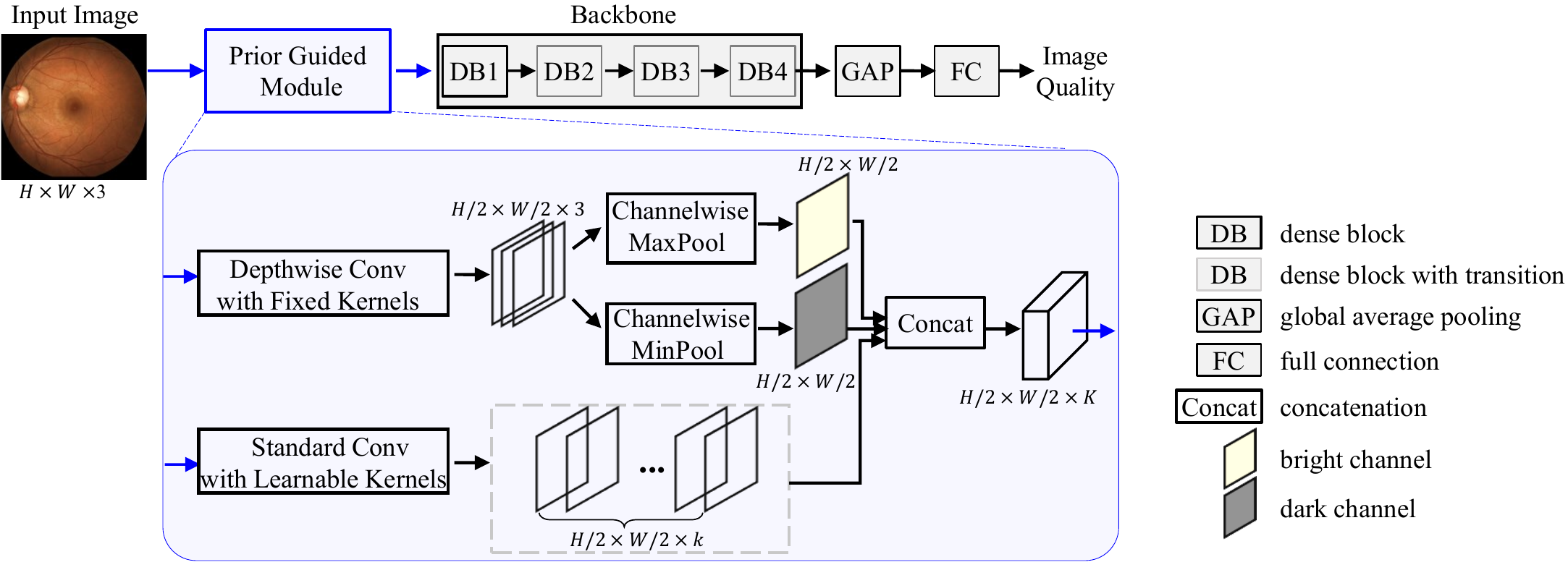}
	\caption{Architecture of proposed dark and bright channel prior guided network (GuidedNet). Our proposed prior guided module takes the colour retinal image size of $H\times W \times 3$ as input and estimates the dark and bright channel priors via depthwise convolution with fixed Gaussian kernel and channel-wise pooling layer. It outputs a group of feature maps size of $\frac{H}{2} \times \frac{W}{2} \times K$, where $K$ is set to 64.}
	\label{fig:NET}
\end{figure*}

In addition, for further analyzing retinal image quality and evaluating the performance of our model, we re-annotate a noval retinal image quality dataset from the released Retinal Fundus Multi-disease Image Dataset (RFMiD) of “Retinal Image Analysis for multi-Disease Detection Challenge” (RIADD challenge) \cite{quellec2020automatic}, named RIQA-RFMiD. In short, the contributions of this paper can be summarized as follows.

\begin{itemize}
\item We introduce dark channel prior and bright channel prior for Retinal-IQA, and propose a novel deep network named GuidedNet, which makes priors of dark channel and bright channel incorporate in CNNs without increasing extra parameters and can be trained end-to-end.

\item We re-annotate a novel dataset called RIQA-RFMiD for retinal image quality assessment, and the annotations will be made publicly.   

\item We demonstrate the effectiveness of the proposed GuidedNet on Eye-Quality \cite{fu2019evaluation} and RIQA-RFMiD, and experimental results show that our GuidedNet achieves state-of-the-art performances.
\end{itemize}

\section{Proposed method}
\label{sec:proposed method}
In this section, we first present the dark and bright channel priors in retinal image. Then we detail the architecture of our proposed GuidedNet.

\textbf{Dark Channel Prior.} Dark channel prior is first proposed to haze removal in \cite{he2010single}. Thereafter it has been proved that the dark channel prior is a suitable image characteristic in distinguishing whether images are polluted by uneven illumination or not and widely applied to blind image deblurring \cite{pan2016blind} and dynamic scene deblurring \cite{cai2020dark}. The dark channel describes the minimum values in an image patch across all colour channels. Formally, according to \cite{he2010single}, in a colour image $I$, the dark pixel at location $x$ is defined as:
\begin{equation}
	\mathrm{D}(I)(x)=\min _{y \in P(x)}\left(\min _{c \in\{r, g, b\}} I^{c}(y)\right)
\end{equation}
where $x$ and $y$ denote pixel locations, $P(x)$ is an image patch centred on $x$, and $ I^{c}$ is intensity map in colour channel $c$. The Fig. \ref{fig:DARK-BRIGHT} shows the dark channel prior maps of three retinal retinal images graded as quality levels of ‘Good’, ‘Usable’ and ‘Reject’. In good quality retinal image, most dark pixels except for optic disc area have low intensities as the image is captured with even illumination, as shown in Fig. \ref{fig:DARK-BRIGHT}(a2). For retinal images captured with slight uneven illumination as shown in Fig. \ref{fig:DARK-BRIGHT}(b1), dark channel prior map exhibits slight uneven intensities as shown in Fig. \ref{fig:DARK-BRIGHT}(b2). For image captured with serious uneven illumination as shown in Fig. \ref{fig:DARK-BRIGHT}(c1), dark pixels of retinal images affected by strong illumination are brighter than those exposed with week illumination, which results in an uneven dark channel prior map, as shown in Fig. \ref{fig:DARK-BRIGHT}(c2). This implies that dark channel prior map is an intuitive quality measure which indicates whether the image is captured with even illumination.

\begin{table*}[ht]
	\centering
	\caption{ Comparison of the proposed method and state-of-the-arts on Eye-Quality \cite{fu2019evaluation}. * indicates that results are taken from the original papers or reproduced by others. }
	\label{table:performances}
	\begin{tabular}{p{4cm}p{2cm}p{2cm}p{2cm}p{2cm}p{2cm}}
		\hline
		Methods & Accuracy & Precision & Recall & F-score & Parameters\\
		\hline
		BRISQUE \cite{mittal2012no}& 0.7692 & 0.7608 & 0.7095 & 0.7112 &-\\
		NBIQA \cite{ou2019novel}& 0.7917 & 0.7641 & 0.7509 & 0.7441 &-\\
		TS-CNN \cite{yan2018two}& 0.7926 & 0.7976 & 0.7446 & 0.7481 & 1.44M \\
		HVS-based algorithm* \cite{wang2015human}& - & 0.7404 & 0.6945 & 0.6991&-\\
		DenseNet121-RGB* \cite{fu2019evaluation} & - & 0.8194 & 0.8114 & 0.8152& 6.96M\\
		DenseNet121-RGB  \cite{fu2019evaluation} & 0.8568 & 0.8481 & 0.8239 & 0.8315& 6.96M\\
		DenseNet121-MCF* \cite{fu2019evaluation} & - & 0.8645 & 0.8497 & 0.8551& 28.26M\\
		DenseNet121-MCF \cite{fu2019evaluation} & 0.8722 & 0.8563 & 0.8482 & 0.8506 & 28.26M\\
		MR-CNN* \cite{raj2020multivariate} & 0.8843 & 0.8697 & 0.8700 & 0.8694 & 101.80M \\
		Combined model* \cite{muddamsetty2020multi} & - & 0.878 & 0.880 & 0.878 & $\sim 29.3$M\\
		\hline
		GuidedNet(ours) & 0.8923 & 0.8863 & 0.8758 & 0.8803 & 6.96M\\
		\hline
	\end{tabular}
\end{table*}

\begin{figure*}[htb]
	\centering
	\includegraphics[width=0.80\linewidth]{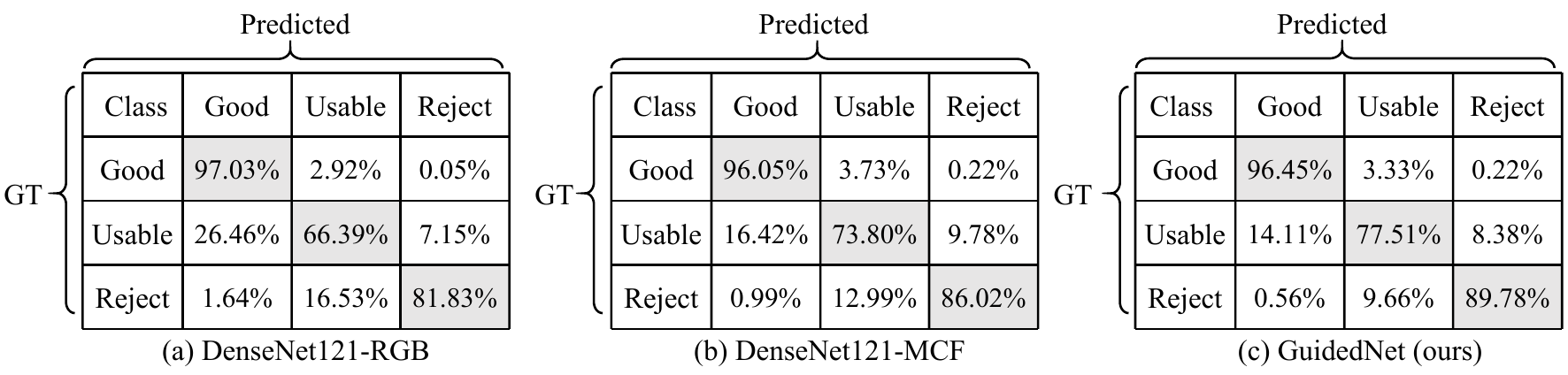}
	\caption{Confusion matrices of DenseNet-121-RGB \cite{fu2019evaluation}, DenseNet-121-MCF \cite{fu2019evaluation} and our GuidedNet on Eye-Quality \cite{fu2019evaluation}.}
	\label{fig:result-confusion}
\end{figure*}

\textbf{Bright Channel Prior.} Bright channel prior describes the maximum values in an image patch across all channels. It is first explored for shadow estimation \cite{panagopoulos2010estimating} and has been widely applied to correct under-exposed images \cite{wang2013automatic}\cite{tao2017low} and image deblurring \cite{pan2016blind}. Here we propose to use bright channel prior to guide the CNN to pay attention to regions with week illumination. Formally, according to \cite{panagopoulos2010estimating}, for an image $I$, the bright pixel at location $x$ is defined as:
\begin{equation}
	\mathrm{B}(I)(x)=\max _{y \in P(x)}\left(\max _{c \in\{r, g, b\}} I^{c}(y)\right)
\end{equation}
where $x$ and $y$ denote pixel locations, $P(x)$ is an image patch centred on $x$, and $ I^{c}$ is intensity map in colour channel $c$. Fig. \ref{fig:DARK-BRIGHT} shows that most pixels are bright in an retina image with good quality, except for some vascular pixels. However, bright pixels in the low-illumination area of retinal images are darker, and  the bright channels present a cloud of dark region. These imply that the bright channel prior is a property to identify abnormal dark region from retinal images.

\textbf{Network Architecture.} The architecture of our GuidedNet is illustrated in Fig. \ref{fig:NET}. It updates from DenseNet-121 \cite{huang2017densely}, which consists of four dense blocks followed by a global average pooling (GAP) layer and a full connection (FC) layer. The FC layer maps the deep features into an image quality level. Different from DenseNet-121 \cite{huang2017densely}, our GuidedNet involves a prior guided module, which replaces the first layer of DenseNet-121 \cite{huang2017densely}. In practice, the dark channel prior is estimated by convolving each colour channel with a fixed Gaussian kernel followed a channel-wise minimisation pooling operator. The bright channel prior can be estimated in a same way by changing the channel-wise minimisation pooling layer to channel-wise maximisation pooling layer. To implement these, we propose to use a depthwise convolutional layer \cite{howard2017mobilenets} with three fixed Gaussian kernels and stride step of 2, followed by a channel-wise MaxPool layer and MinPool layer, as shown in the blue box in Fig. \ref{fig:NET}. In this way, we obtain the bright channel prior map and dark channel prior map both of size $\frac{H}{2}\times \frac{W}{2}$. As the second convolution layer in DenseNet-121 requires input with $K=64$ channels, we learn the rest 62 feature maps via the standard convolution with stride step of 2. Then we concatenate the bright channel prior map, dark channel prior map and learnable feature maps as the output of our prior guided module. We note that the dark and bright channel prior maps involve the forward propagation afterwords and guide the learning of all the parameters in network via back propagation. This enforces the network pay more attention to the informative regions in dark and bright channel prior maps and make correct decisions.

\section{Experimental results}
In this section, we first describe datasets used in this paper, and then detail experimental setting, and finally present experimental results on datasets respectively. Following \cite{fu2019evaluation}, four metrics are employed for evaluation, including average accuracy, precision, recall and F-score.

\subsection{Datasets} 
We validate the effectiveness of the proposed GuidedNet on two retinal image quality datasets. 

1) \textbf{Eye-Quality} \cite{fu2019evaluation}. It contains 28792 retinal images including 12543 images for training and 16249 images for testing. The training dataset contains three-level images: ‘Good’ (8347 images), ‘Usable’ (1876 images) and ‘Reject’ (2320 images), and the testing dataset contains also three-level images: ‘Good’ (8470 images), ‘Usable’ (4559 images) and ‘Reject’ (3220 images). To the best of our knowledge, it is the largest public available dataset with manual quality annotations. The image size ranges from $211 \times 320$ to $3456 \times 5184$ pixels.    


2) \textbf{RIQA-RFMiD}. It is taken from RFMiD of RIADD challenge \cite{pachade2021retinal}. The RFMiD is originally created to develop methods for automatic retinal disease classification of 45 frequent diseases and rare pathopogies. It contains 1920 retinal images for training and 640 for validation. They have a resolution of either $4288 \times 2848$, or $2144 \times 1424$, or $2048 \times 1536$. It is noted that both high-quality and low-quality images are collected to make RFMiD challenging. We ask an expert from The First People’s Hospital of Changde City in China to annotate the quality levels of those images. Among them, 2239 images are labeled as 'Good', 194 images are labeled as 'Usable' and the rest 127 images are labeled as 'Reject'. We name this re-annotated dataset as RIQA-RFMiD. 

\subsection{Experimental setting} 
All the experiments are performed on one GTX1080 Ti GPU, and our GuidedNet is built on the top of implementation of DenseNet-121 \cite{huang2017densely} within the PyTorch framework. For Eye-Quality \cite{fu2019evaluation}, parameters in backbone network are initialized with the pretrained model on ImageNet \cite{deng2009imagenet}. Parameters associated with full connection layers are initialized by Kaiming uniform initialization \cite{he2015delving}. The model is trained by stochastic gradient descent for totally 15 epochs with batch size 8. The initial learning rate is set to 0.01, and then is changed to 0.001 after 10 epoches. For RIQA-RFMiD, parameters are initialized with the corresponding pretrained models based on Eye-Quality \cite{fu2019evaluation}. Parameters are optimized by stochastic gradient descent with learning rate of 0.001. The optimization process is also performed for 15 epochs with batch size of 8. The 5-fold cross-validation is used to validate the models on RIQA-RFMiD.  

Following \cite{fu2019evaluation}, we first crop the field of view (FoV) via Hough Circle Transform, and then pad the short side with zero to make the width and height of the cropped field of view regions be equal length. Finally, we rescale padded regions to $224 \times 224$ pixels. To enrich training data, images are augmented by random vertical and horizontal flipping and rotation.

\begin{table*}[ht]
	\centering
	\caption{Performance comparisons of different settings for batch size and learning rate on Eye-Quality\cite{fu2019evaluation}.}
	\label{table:ablation-parameters}
	\begin{tabular}{p{4.6cm}|p{2.7cm}|p{1.5cm}p{1.5cm}p{1.5cm}p{1.5cm}p{1.2cm}}
		\hline
		Parameters &
		Network & Accuracy & Precision & Recall & F-score & F-std\\
		\hline
		\multirow{2}{*}{b4-lr0.01}& DenseNet121 \cite{huang2017densely} & 0.8825 & 0.8706 & 0.8608 &  0.8638 & 0.0026 \\
		& GuidedNet & 0.8883 & 0.8715 & 0.8747 & 0.8723 & 0.0041\\
		\hline
		\multirow{2}{*}{b4-lr0.01(0.001 after 10 epoches)} & DenseNet121 \cite{huang2017densely} & 0.8834 & 0.8711 & 0.8624 & 0.8641 & 0.0083 \\
		& GuidedNet & 0.8888 & 0.8751 & 0.8710 & 0.8712 & 0.0033 \\
		\hline
		\multirow{2}{*}{b8-lr0.01} & DenseNet121 \cite{huang2017densely} & 0.8874 & 0.8799 & 0.8621 & 0.8694 & 0.0066 \\
		& GuidedNet & 0.8895 & 0.8831 & 0.8622 & 0.8716 & 0.0069\\
		\hline 
		\multirow{2}{*}{b8-lr0.01(0.001 after 10 epoches)} & DenseNet121 \cite{huang2017densely} & 0.8895 & 0.8773 & 0.8700 & 0.8707 & 0.0023 \\
		& GuidedNet & \textbf{0.8923} & \textbf{0.8863} & \textbf{0.8758} & \textbf{0.8803} & \textbf{0.0018} \\
		\hline
	\end{tabular}
\end{table*}

\subsection{Experiments on Eye-Quality}
\label{exp:eye-quality}

\textbf{Comparison with the state-of-the-arts.} On Eye-Quality \cite{fu2019evaluation}, we compare our GuidedNet with eight methods: BRISQUE \cite{mittal2012no}, NBIQA \cite{ou2019novel}, TS-CNN \cite{yan2018two}, HVS-based algorithm \cite{wang2015human}, DenseNet121-RGB \cite{fu2019evaluation}, DenseNet121-MCF \cite{fu2019evaluation}, Multivariate-Regression CNN (MR-CNN) \cite{raj2020multivariate} and Combine model \cite{muddamsetty2020multi}. The first three methods are originally designed for Natural-IQA, in which BRISQUE \cite{mittal2012no} and NBIQA \cite{ou2019novel} use hand-crafted features and TS-CNN \cite{yan2018two} utilizes deep features. Their results are obtained by adjusting authors' codes to Retinal-IQA task with Eye-Quality \cite{fu2019evaluation}. Considering that these three Natural-IQA methods aim to regress natural image quality scores while Retinal-IQA task prefers to classify retinal image into three quality grades, regression should be replaced by classification in the three Natural-IQA methods to make them adapt to Retinal-IQA. Specifically, in the implementation of BRISQUE \cite{mittal2012no} and NBIQA \cite{ou2019novel}, we directly predict retinal image quality label rather than regressing quality score. They are conducted by Matlab2014A on the platform of a PC with an i5 CPU and 16 GB RAM. In the implementation of TS-CNN \cite{yan2018two}, we use a classification layer with three-dimensional outputs to replace the original regression layer with one-dimensional output, and adopt multi-class cross-entropy loss. The last five methods are specially designed for Retinal-IQA. The HVS-based algorithm \cite{wang2015human} uses hand-crafted features and the DenseNet121-RGB \cite{fu2019evaluation}, DenseNet121-MCF \cite{fu2019evaluation}, MR-CNN \cite{raj2020multivariate} adopt deep features, and Combine model \cite{muddamsetty2020multi} ensembles hand-crafted and deep features. The results of HVS-based algorithm \cite{wang2015human}, DenseNet121-RGB \cite{fu2019evaluation} and DenseNet121-MCF \cite{fu2019evaluation} are taken from paper \cite{fu2019evaluation}. For fair comparison, the accuracy results of HVS-based algorithm \cite{wang2015human}, DenseNet121-RGB \cite{fu2019evaluation} and DenseNet121-MCF \cite{fu2019evaluation} are not presented in this paper because they do not report their corrected results. We also provide our reproductions of DenseNet121-RGB \cite{fu2019evaluation} and DenseNet121-MCF \cite{fu2019evaluation}. The results of MR-CNN \cite{raj2020multivariate} and Combine model \cite{muddamsetty2020multi} are directly taken from their original papers.
\begin{figure}[htb]
	\centering
	\includegraphics[width=0.80\linewidth]{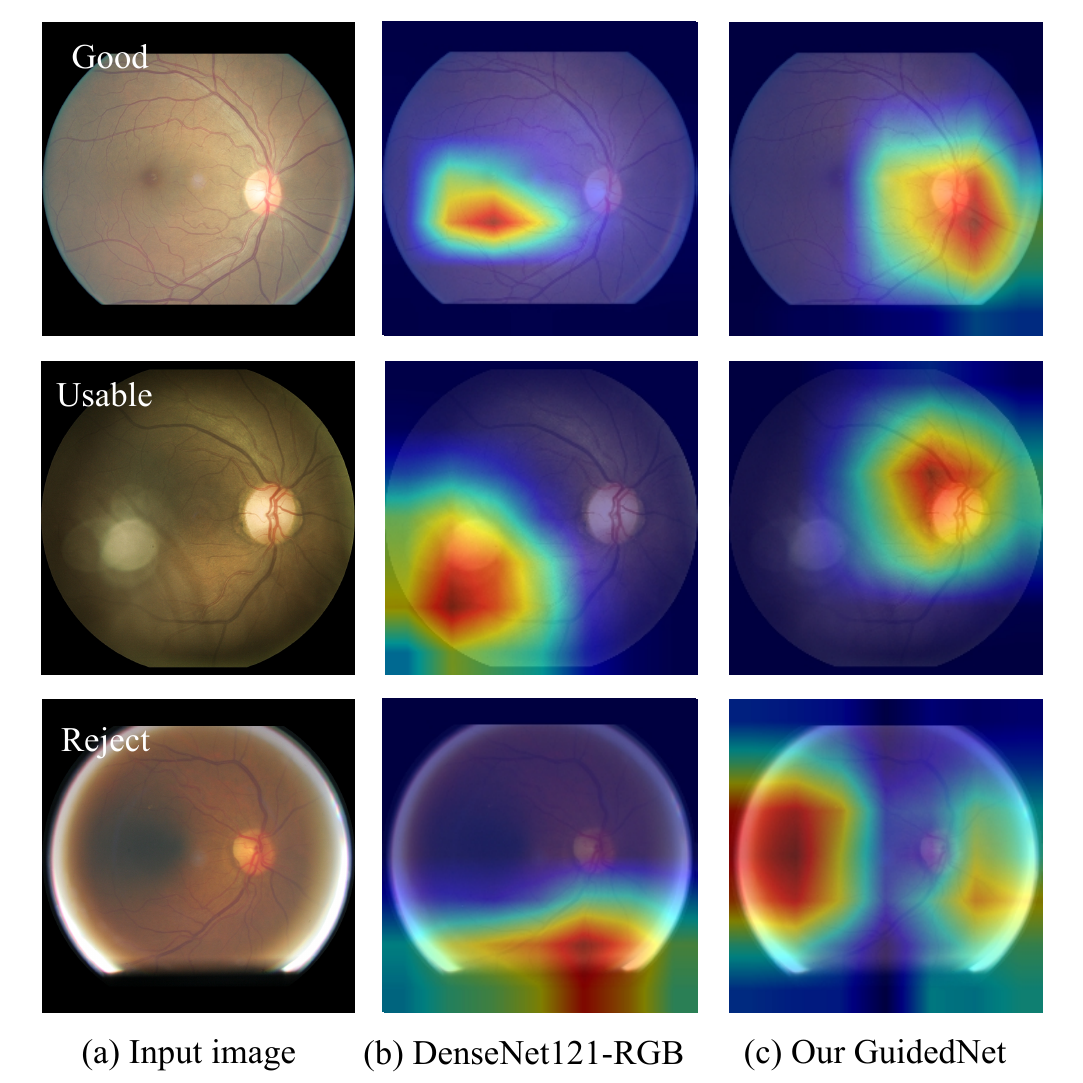}
	\caption{Examples of Grad-cams of different models generated according to \cite{selvaraju2017grad}. These images are wrongly predicted by DenseNet121-RGB \cite{fu2019evaluation}, but correctly identificated by our GuidedNet. Obviously, incorporating dark and bright channel prior, the model prefers to learn discriminate information from informative regions highlighted by prior.}
	\label{fig:cam-v01}
\end{figure}

\begin{table}[ht]
	\centering
	\caption{Effectiveness of dark and bright channel prior on Eye-Quality \cite{fu2019evaluation}. Baseline*: DenseNet121-RGB \cite{fu2019evaluation}. Baseline: DenseNet121 \cite{huang2017densely}. D: Dark channel prior. B: Bright channel prior. DB: Dark and bright channel prior.}
	\label{table:ablation}
	\begin{tabular}{p{1.8cm}p{1cm}p{1cm}p{0.8cm}p{0.8cm}p{0.8cm}}
		\hline
		Methods & Acc & Pre & Rec & F & F-std\\
		\hline
		Baseline* & 0.8568 & 0.8481 & 0.8239 & 0.8315& 0.0085\\
		Baseline & 0.8895 & 0.8773 & 0.8700& 0.8707 & 0.0023\\
		Baseline+D & 0.8916 & 0.8795 & 0.8723 & 0.8740 & 0.0037\\
		Baseline+B  & 0.8936 & 0.8827 & 0.8738 & 0.8767 & 0.0021\\
		Baseline+DB &  \textbf{0.8923} &  \textbf{0.8863} &  \textbf{0.8758} &  \textbf{0.8803} & \textbf{0.0018}\\
		\hline
	\end{tabular}
\end{table}

\begin{table*}[ht]
	\centering
	\caption{Performances of different methods on RIQA-RFMiD.}
	\label{table:performance-isbi2021}
	\begin{tabular}{p{3.3cm}p{1.8cm}p{1.8cm}p{1.8cm}p{1.8cm}p{1.8cm}p{1.5cm}}
		\hline
		Methods & Accuracy & Precision & Recall & F-score & F-std & Parameters\\
		\hline
		DenseNet121 \cite{huang2017densely} & 0.9046 & 0.7011 & 0.6115 & 0.6337 & 0.0103 & 6.96M \\
		DenseNet121-MCF \cite{fu2019evaluation} & 0.9029 & 0.6850 & 0.6477 & 0.6586 & 0.0073 & 28.26M \\
		GuidedNet(ours) &  \textbf{0.9066} &  \textbf{0.6976} & \textbf{0.6574} &  \textbf{0.6613} & \textbf{0.0080} & 6.96M \\
		\hline
	\end{tabular}
\end{table*}

\begin{figure*}[htb]
	\centering
	\includegraphics[width=0.8\linewidth]{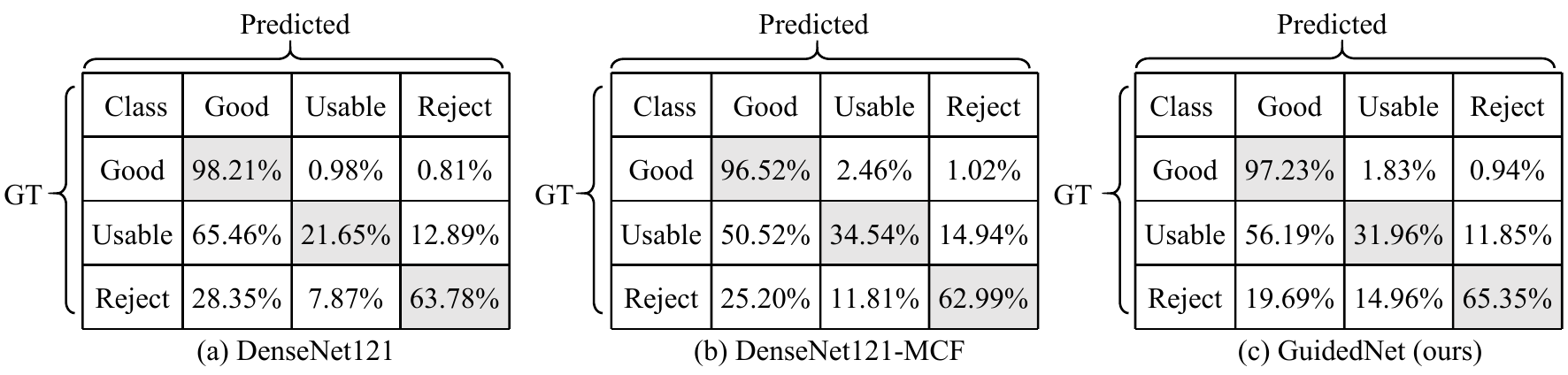}
	\caption{Confusion matrices of DenseNet121 \cite{huang2017densely} and DenseNet121-MCF \cite{fu2019evaluation} and our GuidedNet on RIQA-RFMiD.}
	\label{fig:result-confusion-isbi2021}
\end{figure*} 

We report results in Table \ref{table:performances}, where results marked with * are taken from the original papers or reproduced by others while those without any mark are our reproduction. The implementation of our methods and our reproduction of other methods are conducted three times, and the average results are reported. From Table \ref{table:performances}, we have following observations. Comparing with methods using hand-crafted features, i.e. BRISQUE\cite{mittal2012no}, NBIQA \cite{ou2019novel} and HVS-based algorithm \cite{wang2015human}, methods based on deep features have better performances on Retinal-IQA. For methods using deep features, Retinal-IQA methods significantly outperforms Natural-IQA methods, i.e. TS-CNN \cite{yan2018two}. Compared with DenseNet121-RGB* \cite{fu2019evaluation}, our GuidedNet has a remarkable improvement $6.51\%$ in terms of F-score, achieving $88.03\%$. In terms of accuracy, our GuidedNet achieves $89.23\%$ which is significantly $3.55\%$ higher than DenseNet121-RGB \cite{fu2019evaluation}. As for DenseNet121-MCF* \cite{fu2019evaluation}, the F-score of our GuidedNet outperforms it by $2.97\%$ and the parameters of our GuidedNet are much lighter, approximately 1/4 of DenseNet121-MCF* \cite{fu2019evaluation}. Compared with MR-CNN \cite{raj2020multivariate}, F-score and accuracy of our model are superior by $1.09\%$ and $0.80\%$, respectively. Moreover, the parameters of MR-CNN \cite{raj2020multivariate} are approximately 15 times of our model. Compared with Combined model \cite{muddamsetty2020multi}, our GuidedNet is slightly higher in F-score and our model has less parameters than it. In short, our model achieves superior performances to the state-of-the-arts, and is much lighter than other improved deep model.  

Furthermore, the confusion matrices of DenseNet121-RGB \cite{fu2019evaluation}, DenseNet121-MCF \cite{fu2019evaluation} and our GuidedNet are shown in Fig. \ref{fig:result-confusion}, respectively. For brevity, this paper only presents the confusion matrices of these methods achieving best F-score. From Fig. \ref{fig:result-confusion}, we can observe that the accuracy of our model for recognizing retinal images graded as ‘Good’ has achieved $96.45\%$, which is slightly inferior to DenseNet121-RGB \cite{fu2019evaluation} by $0.58\%$ and superior to DenseNet121-MCF \cite{fu2019evaluation} by $0.40\%$. However, compared with these models, our model obviously performs better in classification on retinal images graded as ‘Usable’ and ‘Reject’ achieving $77.51\%$ and $89.78\%$ in accuracy respectively, which is more important than recognizing retinal images graded as ‘Good’.   

We further present three examples of gradient-weighted class activation maps (Grad-cams) \cite{selvaraju2017grad} of DenseNet121-RGB \cite{fu2019evaluation} and our GuidedNet in Fig. \ref{fig:cam-v01}. These three examples, labeled as ‘Good’, ‘Usable’ and ‘Reject’ from top to bottom in Fig. \ref{fig:cam-v01}, are wrongly graded as ‘Usable’, ‘Reject’ and ‘Usable’ by DenseNet121-RGB \cite{fu2019evaluation} but correctly classified by our GuidedNet. As shown in the second column in Fig. \ref{fig:cam-v01}, DenseNet121-RGB \cite{fu2019evaluation} fails to capture the most informative information for quality prediction. However, our GuidedNet tends to learn more discriminate information from regions highlighted by prior. As shown in the third column in Fig. \ref{fig:cam-v01}, our GuidedNet focuses on the distorted regions with strong and weak illumination for the classification of ‘Reject’ images, while it prefers to learn information from optic disc for the classification of ‘Good’ and ‘Usable’ images.   


\textbf{Effectiveness of dark and bright channel prior.} To validate the effects of dark and bright channel prior, we perform ablation study on Eye-Quality \cite{fu2019evaluation}. The average results are reported in Table \ref{table:ablation}, where the method marked with * is trained following settings of \cite{fu2019evaluation} while methods without any marks follow settings of this paper. In addition, the baseline in Table \ref{table:ablation} refers to DenseNet121 \cite{huang2017densely}. From Table \ref{table:ablation}, we can see that (1) the result of baseline using our settings are superior than that using settings of \cite{fu2019evaluation}. The possible reason is that random drifting in \cite{fu2019evaluation} may cause label noise, because distortions in retinal images are uneven. (2) Our GuidedNet achieves best overall evaluation value F-score while incorporating both dark and bright prior into network. (3) Embedding either dark or bright prior can also contribute to improve the performance of Retinal-IQA.

\begin{table*}[ht]
	\centering
	\caption{Influences of initial weights on RIQA-RFMiD.}
	\label{table:ablation-pretrained}
	\begin{tabular}{p{4.2cm}|p{2.6cm}|p{1.5cm}p{1.5cm}p{1.5cm}p{1.5cm}p{1.5cm}}
		\hline
		Initial weights & Network  & Accuracy & Precision & Recall & F-score & F-std\\
		\hline
		\multirow{2}{*}{pretrained on Imagenet \cite{deng2009imagenet}} & DenseNet121 \cite{huang2017densely} & 0.8924 & 0.6372 & \textbf{0.6289} & 0.6051 & \textbf{0.0084}\\
		& GuidedNet & \textbf{0.8967} & \textbf{0.6580} & 0.6264 & \textbf{0.6107} & 0.0154 \\
		\hline
		\multirow{2}{*}{pretrained on Eye-Quality \cite{fu2019evaluation}} & DenseNet121 \cite{huang2017densely} & 0.9046 & \textbf{0.7011} & 0.6115 & 0.6337 & 0.0103 \\
		& GuidedNet & \textbf{0.9066} & 0.6976 & \textbf{0.6574} & \textbf{0.6613} & \textbf{0.0080}\\
		\hline
	\end{tabular}
\end{table*}

\textbf{Ablation study.} We look into the influences of different settings, i.e. batch size and learning rate. Table \ref{table:ablation-parameters} reports the performances of four settings of two networks including DenseNet121 \cite{huang2017densely} and our GuidedNet. In Table \ref{table:ablation-parameters}, four settings are seperately defined as: 1) b4-lr0.01, means that models are trained 20 epoches with bacth size 4 and learning rate 0.01. 2) b4-lr0.01(0.001 after 10 epoches), means that models are first trained 10 epoches with bacth size 4 and learning rate 0.01 and then trained another 10 epoches with decreased learning rate 0.001. 3) b8-lr0.01, means that models are optimized for 15 epoches with batch size 8 and learning rate 0.01. 4) b8-lr0.01(0.001 after 10 epoches), means that models are trained totally 15 epoches and the initial learning rate is set to 0.01, and the learning rate is decreased to 0.001 after 10 epoches. From Table \ref{table:ablation-parameters}, we can observe that (1) our GuidedNet has achieved the best performance when parameters are set as b8-lr0.01(0.001 after 10 epoches). (2) with the same setting, our GuidedNet incorporating dark and bright channel prior is superior to DenseNet121 \cite{huang2017densely}. 


\subsection{Experiments on RIQA-RFMiD}
\label{exp:RIQA-RFMiD} 


\textbf{Comparison with the state-of-the-arts.} On RIQA-RFMiD, we compare our GuidedNet with DenseNet121 \cite{huang2017densely} and DenseNet121-MCF \cite{fu2019evaluation}. We run three times for each model, and report their average results in Table \ref{table:performance-isbi2021}. From Table \ref{table:performance-isbi2021}, it is obvious that (1) our model has the best performance in both F-score and accuracy. (2) Compared with DenseNet121 \cite{huang2017densely}, our GuidedNet have clear performance gains $2.76\%$ in F-score with the same parameter scales. (3) Compared with DenseNet121-MCF \cite{fu2019evaluation}, our GuidedNet slightlt outperforms it but the parameters of GuidedNet are approximately 1/4 of DenseNet121-MCF \cite{fu2019evaluation}. 

We also report the confusion matrices of DenseNet121 \cite{huang2017densely}, DenseNet121-MCF \cite{fu2019evaluation} and our GuidedNet in Fig.\ref{fig:result-confusion-isbi2021}. From Fig.\ref{fig:result-confusion-isbi2021}, it is noted that our GuidedNet achieves considerable accuracy of $97.23\%$ in classifying ‘Good’ images, which is slightly inferior to DenseNet121 \cite{huang2017densely}. But DenseNet121 \cite{huang2017densely} is more likely to mistakely classify ‘Usable’ and ‘Reject’ images into ‘Good’ than our GuidedNet. Compared with DenseNet121-MCF \cite{fu2019evaluation}, our GuidedNet has higher accuracy in the classification of ‘Good’ and ‘Reject’ images. In addition, our GuidedNet is inferior to DenseNet121-MCF \cite{fu2019evaluation} in the identification of ‘Usable’ images. However, it has lower error rate $11.85\%$ of classifying these ‘Usable’ images into ‘Reject’ level than DenseNet121-MCF \cite{fu2019evaluation}, which makes more sense than distinguishing ‘Usable’ and ‘Good’ retinal images clinically.

\textbf{Influences of initial weights.}
To study the perfomance comparison of initial weights on RIQA-RFMiD, we compare two initialization ways. One way is the weights of base network are initialized from pretained model based on Imagenet \cite{deng2009imagenet}, and the weights of full connection layers are initialized randomly. Another way is loading initial weights from pretained models based on Eye-Quality \cite{fu2019evaluation}. For each initialization way, we validate on DenseNet121 \cite{huang2017densely} and our GuidedNet. The performances are reported in Tabel \ref{table:ablation-pretrained}. From Tabel \ref{table:ablation-pretrained}, it is obvious that our GuidedNet always outperforms DenseNet121 \cite{huang2017densely} under the same initialization, which is caused by incorporated dark and bright channel prior. In addition, networks initialized by pretrained models based on Eye-Quality \cite{fu2019evaluation} are superior to networks initialized by pretrained model based on Imagenet \cite{deng2009imagenet}. The reason may be the pretrained model based on Imagenet \cite{deng2009imagenet} is trained for object classification, while pretrained models based on Eye-Quality \cite{fu2019evaluation} have been optimized for Retinal-IQA, which have captured much retinal image quality information.

\section{Conclusion}
\label{sec:conclusion}
This paper presents a simple framework named GuideNet for retina image quality assessment,and re-creater a new retinal imgae quality dataset called RIQA-RFMiD. It introduces dark and bright channel priors to predict image quality. The proposed GuidedNet builds a dark and bright channel prior guided layer to highlight image quality prior and does not increase much model burden. Our model does not require auxiliary landmark detection module and can be trained end-to-end. Validations on Eye-Quality \cite{fu2019evaluation} and our RIQA-RFMiD show the superior performances of our GuidedNet and the effectiveness of dark and bright channel priors in retinal image quality assessment.

\section*{Acknowledgments}
The authors would like to thank Dr.~Decheng Yu from The First People's Hospital of Changde City for annotating RIQA-RFMiD. Prof.~Beiji Zou is partially supported by the National Key R\&D Program of China (NO. 2018AAA0102100), and Dr.~Qing Liu is partially supported by the National Natural Science Foundation of China (No. 62006249) and Changsha Municipal Natural Science Foundation (kq2014135). 





\bibliographystyle{IEEEbib}
\bibliography{refs}




\end{document}